**This is the accepted manuscript (postprint) of the following article:**
M. Moradi, E. Salahinejad, E. Sharifi, L. Tayebi, *Controlled drug delivery from chitosan-coated heparin-loaded nanopores anodically grown on nitinol shape-memory alloy*, Carbohydrate polymers, 314 (2023) 120961.
https://doi.org/10.1016/j.carbpol.2023.120961

# Controlled drug delivery from chitosan-coated heparin-loaded nanopores anodically grown on nitinol shape-memory alloy

M.R. Moradi [a], E. Salahinejad [*, a], E. Sharifi [b], L. Tayebi [c]

[a] Faculty of Materials Science and Engineering, K. N. Toosi University of Technology, Tehran, Iran

[b] Department of Tissue Engineering and Biomaterials, School of Advanced Medical Sciences and Technologies, Hamadan University of Medical Sciences, Hamadan, Iran

[c] Marquette University School of Dentistry, Milwaukee, WI 53233, USA

## Abstract

Nitinol (NiTi shape-memory alloy) is an interesting candidate in various medical applications like dental, orthopedic, and cardiovascular devices, owing to its unique mechanical behaviors and proper biocompatibility. The aim of this work is the local controlled delivery of a cardiovascular drug, heparin, loaded onto nitinol treated by electrochemical anodizing and chitosan coating. In this regard, the structure, wettability, drug release kinetics, and cell cytocompatibility of the specimens were analyzed *in vitro*. The two-stage anodizing process successfully developed a regular nanoporous layer of Ni-Ti-O on nitinol, which considerably decreased the sessile water contact angle and induced hydrophilicity. The application of the chitosan coatings controlled the release of heparin mainly by a diffusional mechanism, where the drug release mechanisms were evaluated by the Higuchi, first-order, zero-order, and Korsmeyer-Pepass models. Human umbilical cord endothelial cells (HUVECs) viability assay also showed the non-cytotoxicity of the samples, so that the best performance was found for the chitosan-coated samples. It is concluded that the designed drug delivery systems are promising for cardiovascular, particularly stent applications.

* Corresponding Author: Email Address: <salahinejad@kntu.ac.ir>

## 1. Introduction

Coronary diseases originate from the formation of plaque on arteries' walls and are one of the most serious health problems as they can lead to heart attacks. Due to shape memory behaviors of nitinol (NiTi alloy), using nitinol stents is a suitable treatment of arterial blockage (Mohammadi et al., 2019a). In this regard, stent thrombosis is the most typical complication experienced by patients after using a stent. The delay in the endothelium repair process is another problem of stents (Liu et al., 2014a). Approaches like the deposition of coatings and/or immobilization of biomolecules can alter the endothelium repair process somewhat. But these surface engineering methods cannot entirely prevent the thrombus development and address the endothelium repair process (Gong et al., 2019a).

Introduction of drug-eluting stents was a revolution in treating stent-related diseases, where immunosuppressive drugs like Sirolimus (Thipparaboina et al., 2013) and anti-cancer drugs such as Paclitaxel (Bhargava et al., 2006) have been incorporated in these stents. In addition, Dipyridamole (Thipparaboina et al., 2013), Valsartan (Peters et al., 2009), Leflunomide (Deuse et al., 2008), Succinobucol (Watt et al., 2013), Biolimus A9 (Tada et al., 2010), Everolimus (Mani et al., 2008), Tacrolimus (Bartorelli et al., 2003), and Heparin (Shen et al., 2022.) are some blood-modifying drugs that have been used in stents so far. Among them, heparin as a highly sulfated glycosaminoglycan reacts with an extensive range of proteins and growth factors, which are involved in a lot of bioprocesses like blood coagulation, immunology, cell growth, pathological, and physiological processes. Heparin is also known to inhibit smooth muscle cells (SMC) adhesion and proliferation and increase endothelial cell proliferation (Yu et al., 2021.). Nevertheless, a controlled supply of heparin is required to benefit from all of its features and to prevent any negative responses, suggesting the need to employ controlled drug delivery systems.

Anodically grown nanostructures on nitinol and titanium alloys in the forms of nanotubes and nanopores are promising as a platform for drug delivery and could also improve some bio-performances like bioactivity. Considerable mechanical properties, corrosion resistance, specific surface area, and the ability to control the thickness and diameter of nanotubes are the unique features of nanotubular structures (W. Chen et al., 2015; Davoodian et al., 2020; Jarosz et al., 2016; Meng et al., 2009; Z. Yang et al., 2012; Yao et al., 2014). Despite these significant advantages, Ni-Ti-O nanotubes grown on nitinol have a thickness limit not exceeding 1 μm. But the thickness of Ni-Ti-O nanoporous structures can reach up to 160 μm (Mousavi et al., 2021), which provides a high drug loading capacity, where the thickness of 11 μm has exhibited the highest cell biocompatibility (Hang et al., 2018). These nanostructures also improve the growth of endothelial cells, reduce the growth of muscle cells, and decrease the number of platelets attached to the surface beneficially (Gong et al., 2019b; Q. Huang et al., 2017).

Various studies have been done on drug loading and release from Ni-Ti-O nanotubes (Davoodian et al., 2020; Mohammadi et al., 2019b). Also, the drug delivery potential of Ti-O nanoporous structures grown on titanium alloys loaded with ibuprofen and gentamicin has been verified (Jarosz et al., 2016). But both drug-loaded nanotubular and nanoporous structures suffer from an initial burst release of the incorporated therapeutic agents, which can cause toxicity and inefficient therapy (Davoodian et al., 2020; Jarosz et al., 2016). To control the kinetics of drug delivery from these nanostructures, using biopolymeric coatings is one of the most proper methods (W. Chen et al., 2015; Davoodian et al., 2020; Gulati et al., 2012). In this regard, chitosan and poly(lactic-co-glycolic acid) (PLGA) are the most commonly utilized polymers due to their proper biocompatibility and biodegradability (Davoodian et al., 2020). Chitosan is a natural, non-human origin, cationic polysaccharide with moderate ionic strength. Accelerating the re-endothelialization and helping to reduce the growth of SMCs are other advantages of chitosan (J. Chen et al., 2016a; Meng et al., 2009; Yao et al., 2014), making it promising for cardiovascular applications.

To our knowledge, there are no reports in the literature on the response of endothelial cells to Ni-Ti-O nanoporous structures, as well as the drug loading and delivery behavior of these nanostructures with and without any polymeric capsulation. Accordingly, the hypothesis of this work is that nanopores grown on nitinol by a modified, two-step anodizing process act as a desired platform for the controlled delivery of heparin, so that a further controlled behavior within the therapeutic window is obtained after coating with chitosan.

## 2. Experimental

### *2.1. Materials*

Nitinol discs (Kellog's Research Labs) with the diameter of 8 mm were used as the substrate. Double-distilled water was utilized to prepare the anodization electrolyte, drug solution, polymer solutions, and phosphate-buffered saline (PBS). Ethylene glycol (Merck, Germany, >99.5%) and sodium chloride (Merck, Germany, >99.5%) were employed to produce the anodization electrolyte. Unfractionated heparin salt from porcine intestinal mucosa (Alfa Aesar, molecular weight: 8-25 kDa, IU>= 100/mg) and chitosan (Sigma Aldrich, molecular weight: 140469.4 g/mol, acetylation degree: 84.2%) were used as the anticoagulant drug and encapsulating polymer, respectively.

### *2.2. Specimen preparation*

#### *2.2.1. Anodization*

First, the nitinol samples were polished by sandpapers numbered from 240 to 1000. Then, using a copper glue, a piece of wire was attached to the back of the samples, and an aquarium glue was used to seal the wire connection for the anodizing process. After 24 h of the aquarium glue drying, the samples were polished to 2500 grit to achieve mirror-like surfaces. Finally, the specimens were washed in an ultrasonic bath (Parsinic 7500s) in water and ethanol for 10 min (MP-NiTi sample).

For the anodization of the polished samples, 100 ml of a solution of 95 % ethylene glycol, 5 % water, and 0.3 M NaCl was used. The anodizing process was done at 10 V at ambient temperature for 60 min, while a platinum cathode was used. To obtain a regular nanostructured layer, the anodizing process consisted of two static and dynamic stages; the first static mode of 30 min and the following dynamic step of 30 min, according to Ref. (Mousavi et al., 2021). Finally, the specimens were placed at 60°C for 60 min for drying (NP-NiTi sample).

#### *2.2.2. Drug loading*

A heparin solution of 5 mg/ml was used for loading the drug inside Ni-Ti-O nanopores by a vacuum set-up (Figure 1). For this purpose, 10 ml of the heparin solution was first prepared. The sample was placed inside a round-bottomed balloon and after placing a vacuum valve on the balloon, the valve was connected to a pump. The setup was placed under vacuum for 15 min, and the drug solution was injected into the setup using a syringe. The duration of the loading process was 24 h. Afterwards, the sample was removed and placed outside for 24 h air drying (NP-Hep sample).

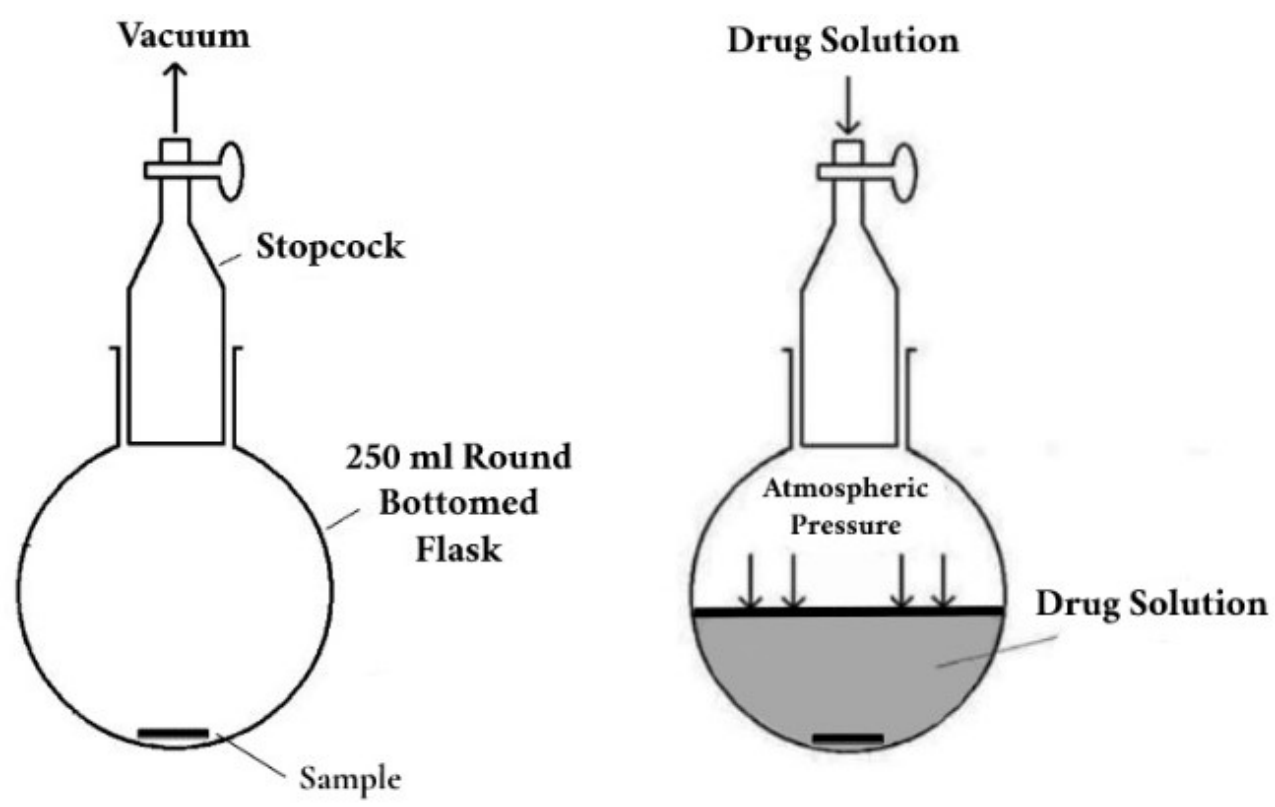

Figure 1. Schematic illustration of the drug loading setup, adapted from Ref. *(Byrne & Deasy, 2002)*.

### *2.2.3. Encapsulation with chitosan coatings*

Chitosan was dissolved in 0.2 % acetic acid solution at 0.1 and 0.2 wt% to apply chitosan coatings with two thicknesses. After preparing 10 ml of the solutions, the NP-Hep samples were placed in the solution for 10 sec and then in an oven at 50°C for 15 min for drying (0.1 and 0.2 wt% Chi samples)

## ***2.3. Structural studies***

The morphology and cross section of the specimens were imaged by a field emission scanning electron microscope (FESEM, TESCAN-XMU, Mira3) with an accelerating voltage of 15 kV. The mean pore diameter of the specimens was also determined by the ImageJ software.

## ***2.4. Wettability analysis***

To determine the sessile water contact angle of the surfaces, a drop of water (10 µl) was dripped and then photographed from an angle perpendicular to the cross-section of the samples. Afterwards, contact angles were measured using the ImageJ software. The analyses were repeated at least three times, and the reported values are the average of these values.

### *2.5. Heparin release kinetics studies*

To assess the release kinetics of heparin from the samples, the NP-Hep, 0.1 and 0.2 wt% Chi samples were placed in 10 ml of the PBS solution. After 3, 6, 9, 12, 24, 48, 72, 96, 144, and 168 h of immersion at 37 °C, 3 ml of the medium was taken and then the same amount of the fresh medium was replaced. The Toluidine Blue O (TBO) assessment was used to evaluate heparin release, based on Ref. (J. Chen et al., 2016b). The taken solutions were analyzed to estimate the heparin content by an ultraviolet-visible spectrometer (Perkin Elmer) at the wavelength of 631 nm. To draw the calibration graph, heparin concentrations of 250, 500, 750, 1000, 1500, and 1700 µg/ml were used. To further investigate the kinetics of heparin release, the Higuchi, first-order, zero-order, and Korsmeyer-Pepass models were fitted to the experimental results.

### *2.6. Cytotoxicity assessment*

Human umbilical cord endothelial cells (HUVECs) obtained from the National Cell Bank of Iran were used in this investigation for biocompatibility assessments. The prepared specimens were sterilized in ethanol 70% and afterwards washed with the PBS solution. 200 µL suspension of approximately $1.0 \times 10^4$ HUVECs in the Dulbecco's modified eagle medium containing 20% fetal bovine serum and 1% antibiotics (penicillin/ streptomycin) was transferred onto the specimens located in a 48-well cell culture plate in an incubator with 5% $CO_2$ at 37 °C for 1 and 3 days. A cell counting kit-8 (CCK-8) was used to measure the cell viability with three replicates, based on Ref. (Rahimipour et al., 2020a). Briefly, 20 µl of the CCK-8 solution was added to each well, and after 4 h of incubation, 100 µl of the each well content was transferred to a 96-well microplate. The solution optical density at 450 nm was measured by a ChroMate-4300 microplate reader. One-way analysis of variance was used for statistical evaluations using $P>0.05$.

## 3. Results and discussion

### *3.1. Structural studies*

The FESEM graphs of the NP-NiTi sample are indicated in Figure 2. As can be observed in Figure 2a and b, a homogeneous nanoporous structure has been formed as a result of the utilized two-step anodizing process. According to studies, static anodizing processes normally create an irregular layer on surfaces (Zhao et al., 2019). But the dynamic stage of the process causes the removal of the irregular layer, albeit accompanied by decreasing the thickness of the oxide layer. The reason for this phenomenon is the stimulation of corrosive and oxidizing ions due to stirring, overcoming the surface dissolution over the bottom dissolution (Mousavi et al., 2021). Based on the cross-sectional images (Figure 2c and d), the average thickness of the porous layer was calculated to be about 10 µm which is near the optimal value from the biocompatibility viewpoint (Hang et al., 2018).

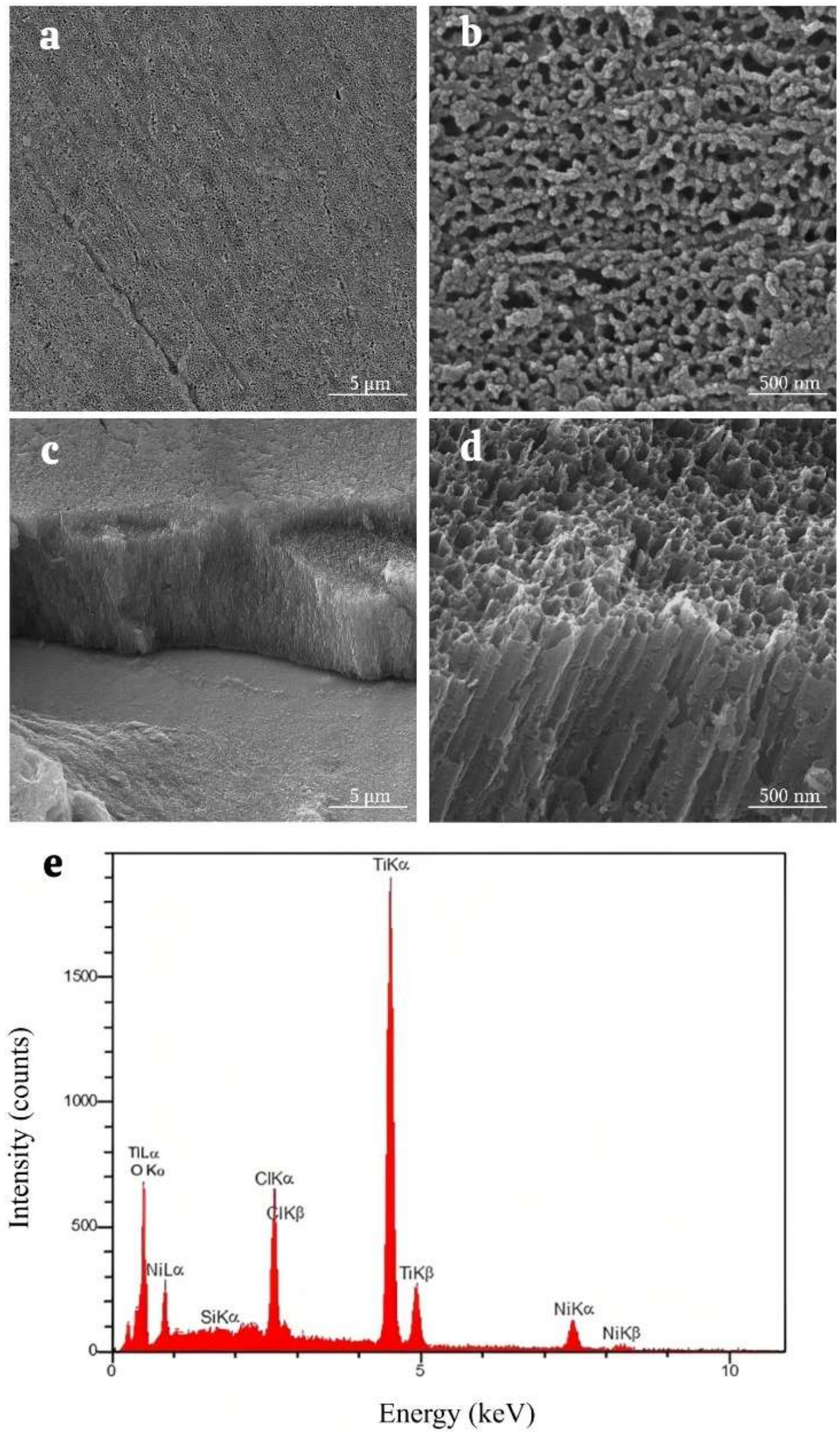

Figure 2. FESEM images of the anodized sample at different magnifications: 10 kx (a) and 100 kx (b) (top-views), and 7 kx (c) and 100 kx (d) (cross-sections), and related EDS pattern (e).

According to energy-dispersive X-ray spectroscopy (EDS) on the anodized sample (Figure 2e), the oxide nature of the nanoporous layer is confirmed as it contains titanium, nickel, oxygen, chlorine, and silicon elements. The presence of silicon is related to the silicon glue used to bond the samples for the anodizing process. According to studies (Hang et al., 2017), Ni-Ti-O nanoporous layers grown by anodization in electrolytes containing NaCl mainly contain amorphous $TiO_2$ and a little amount of NiO. The growth mechanism of nanopores is defined by a balance between the oxidation of the main constitutes of nitinol (nickel and titanium) and the dissolution of the formed oxides that are attacked by Cl ions (Hang et al., 2018). At the beginning of the process, a relatively dense NiO and $TiO_2$ layer is formed based on the following reactions (Hang et al., 2017.; Hang, Zhao, et al., 2017; Mousavi et al., 2021):

$$Ti + 2H_2O \rightarrow TiO_2 + 4H^+ + 4e^- \quad (1)$$

$$Ni + H_2O \rightarrow NiO + 2H^+ + 2e^- \quad (2)$$

By progression of the process, chloride ions with the help of the electric field react with the oxide layer according to the following reactions with the aim of forming nanopores:

$$TiO_2 + 4H^+ + 6Cl^- \rightarrow [TiCl_6]^{2-} + 2H_2O \quad (3)$$

$$NiO + 2H^+ + 4Cl^- \rightarrow [NiCl_4]^{2-} + H_2O \quad (4)$$

To determine the pore size distribution, three images were considered using the ImageJ software from random parts of top-views at the 50 kx magnification. From this analysis (Figure 3), the mean diameter of nanopores was measured to be almost 45 nm. In addition, almost 32% of nanopores are in the range of 35-45 nm, 24% in the range of 45-55 nm, and 23% in the range of 25-35 nm. According to Zhiaho et al. (Gong et al., 2019b), the most optimal diameter of nanotubes for endothelial cells adhesion, migration, and differentiation is 15-30 nm. When the diameter reaches 100 nm, the cell metabolism hardly occurs, causing the cell death. Also, anodic Ni-Ti-O nanoporous structures show better corrosion resistance and cytocompatibility than bare nitinol (Hang et al., 2017).

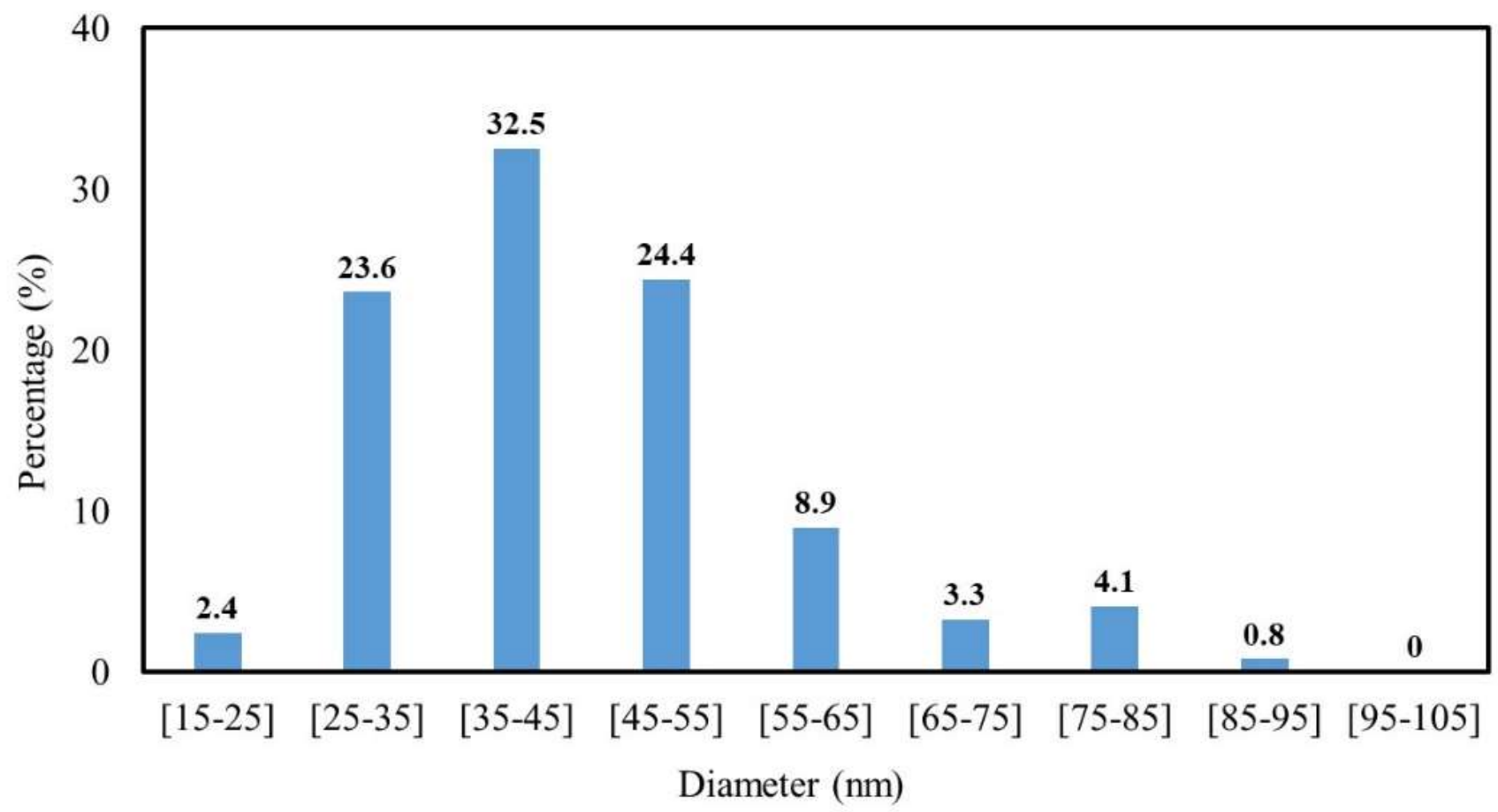

Figure 2. Size distribution of nanopores grown anodically on nitinol.

The FESEM images of the chitosan-coated specimens are also depicted in Figure 4. Using the two chitosan concentrations (0.1 and 0.2 wt%) causes the formation of two thin and thick layers of chitosan on the anodically-grown nanoporous layer. The thickness of the chitosan layer obtained at the concentration of 0.1 wt% is about 350 nm, whereas this value is 470 nm for the concentration of 0.2 wt%.

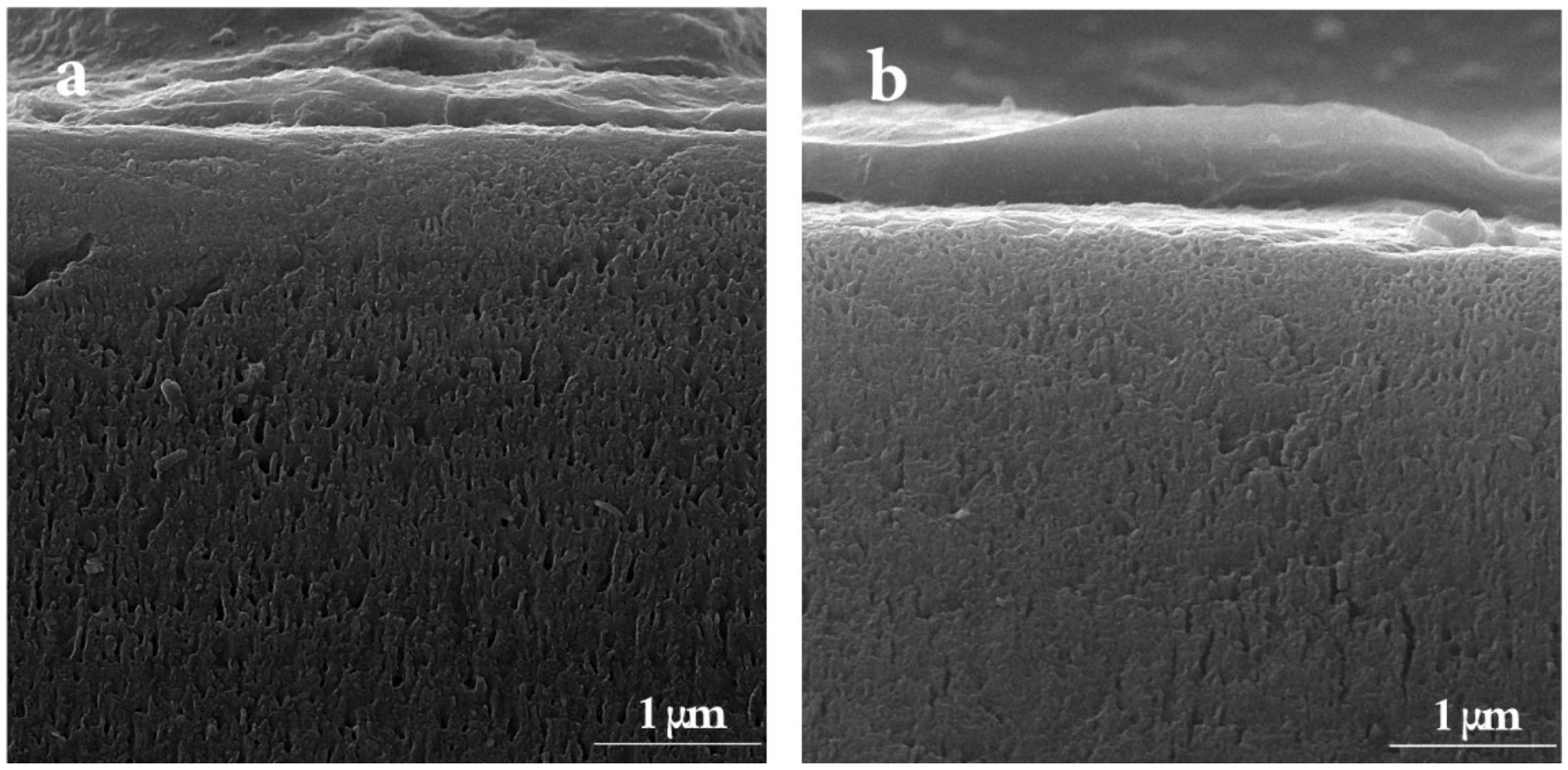

Figure 4. Cross-sectional FESEM images of the 0.1 wt% Chi (a) and 0.2 wt% Chi (b) samples.

### *3.2. Wettability studies*

The wettability of implants affects their bioactivity and biocompatibility, particularly the adhesion of proteins, platelets, and cells (Drugacz et al., 1995; Gong et al., 2019a; Oh et al., 2005). Due to the adsorption of high amounts of water molecules, hydrophilic surfaces can reduce the adsorption of non-specific proteins, decrease the absorption of platelets, and enhance blood compatibility (Gong et al., 2019a; Pan et al., 2016). Also, hydrophilic surfaces can bond with proteins through electrostatic attractions or hydrogen bonds, providing a high density of available molecules attached to the surface for cell attachment and growth and ultimately improving cell activity. Heparin, as the anticoagulant used in this study, is completely hydrophilic and soluble in water and aqueous solvents as it is rich in hydrophilic groups like hydroxyl, carboxyl, amine, and sulfur. Therefore, it is quite logical that it is very easy to be loaded into hydrophilic matrices.

Figures 5 and 6 show the water contact angle testing results on the various specimens. The anodizing process giving rise to the formation of the Ni-Ti-O nanoporous layer causes a significant reduction in the contact angle. The wettability increase after the anodizing process is owing to hydroxylation (development of Ti-OH groups) and surface roughness (Hang et al., 2012). Regarding the other samples, i.e., NP-Hep, 0.1 and 0.2 wt % Chi, the contact angles are lower than MP-NiTi, but higher than NP-NiTi. The higher contact angle of the NP-Hep sample can be owing to the filling of the pores by heparin and the decrease of the surface roughness (J. Huang et al., 2014). Via the same argument, the higher contact angle of the 0.1 and 0.2 wt% Chi samples can be explained.

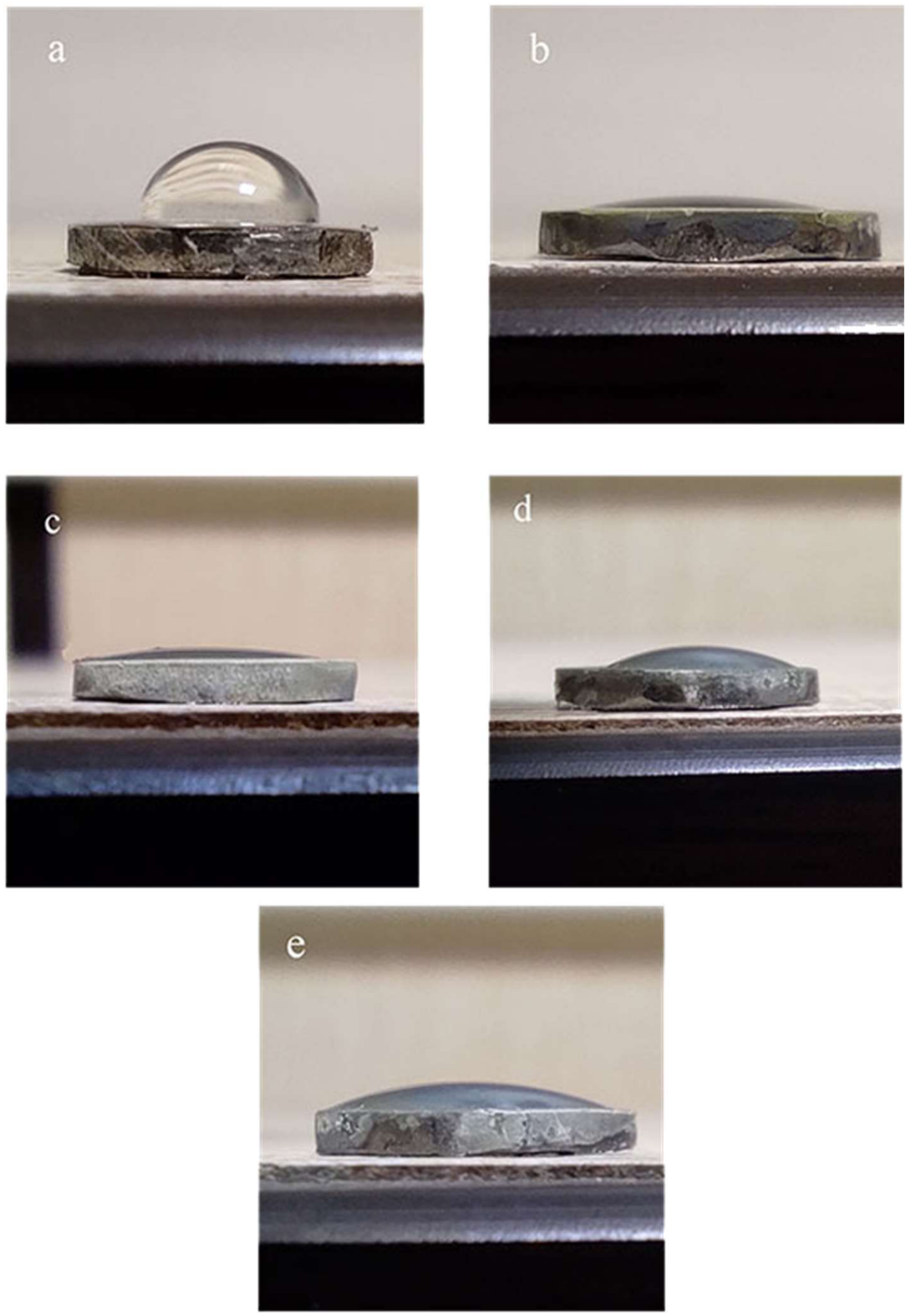

Figure 3. Water droplets on the samples: MP-NiTi (a), NP-NiTi (b), NP-Hep (c), 0.1 wt% Chi (d), and 0.2 wt% Chi (e).

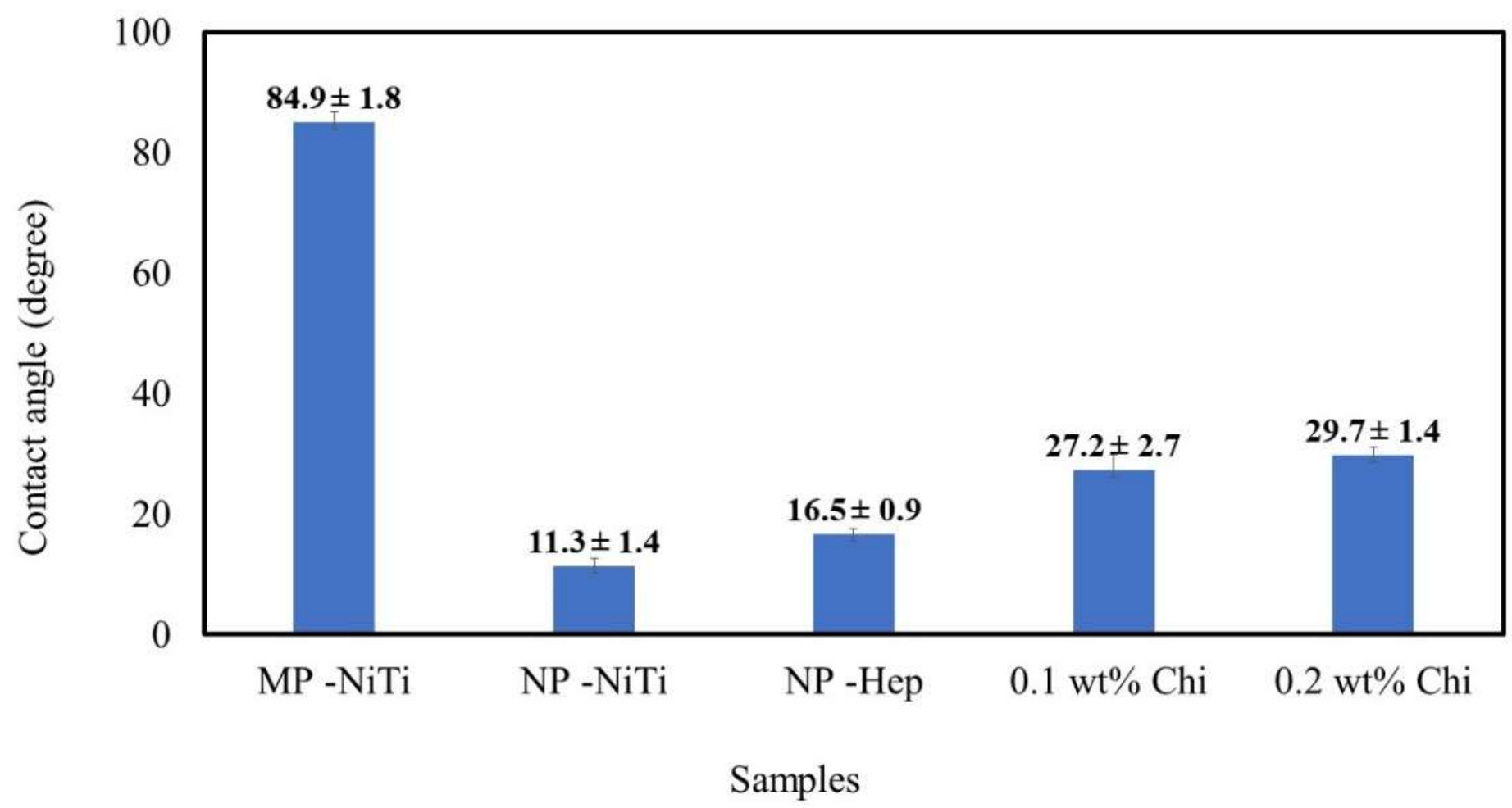

Figure 6. Water contact angle on the different samples.

### *3.3. Drug delivery studies*

Figure 7 demonstrates the behavior of heparin release from the drug-loaded specimens. According to the curves, all of the samples show a two-stage behavior; an initial burst release to the first 12 h and a following sustained mode. About 70 % of heparin is released from the NP-Hep sample during the first 12 h, which is regarded as a rapid and explosive release. The initial rapid release can be owing to the high concentration gradient of heparin at the interface between PBS and nanopores, so that heparin molecules attached to the surface and opening of nanopores are released at this phase. The subsequent sustained release phase is attributed to heparin molecules that have penetrated into the depth of nanopores.

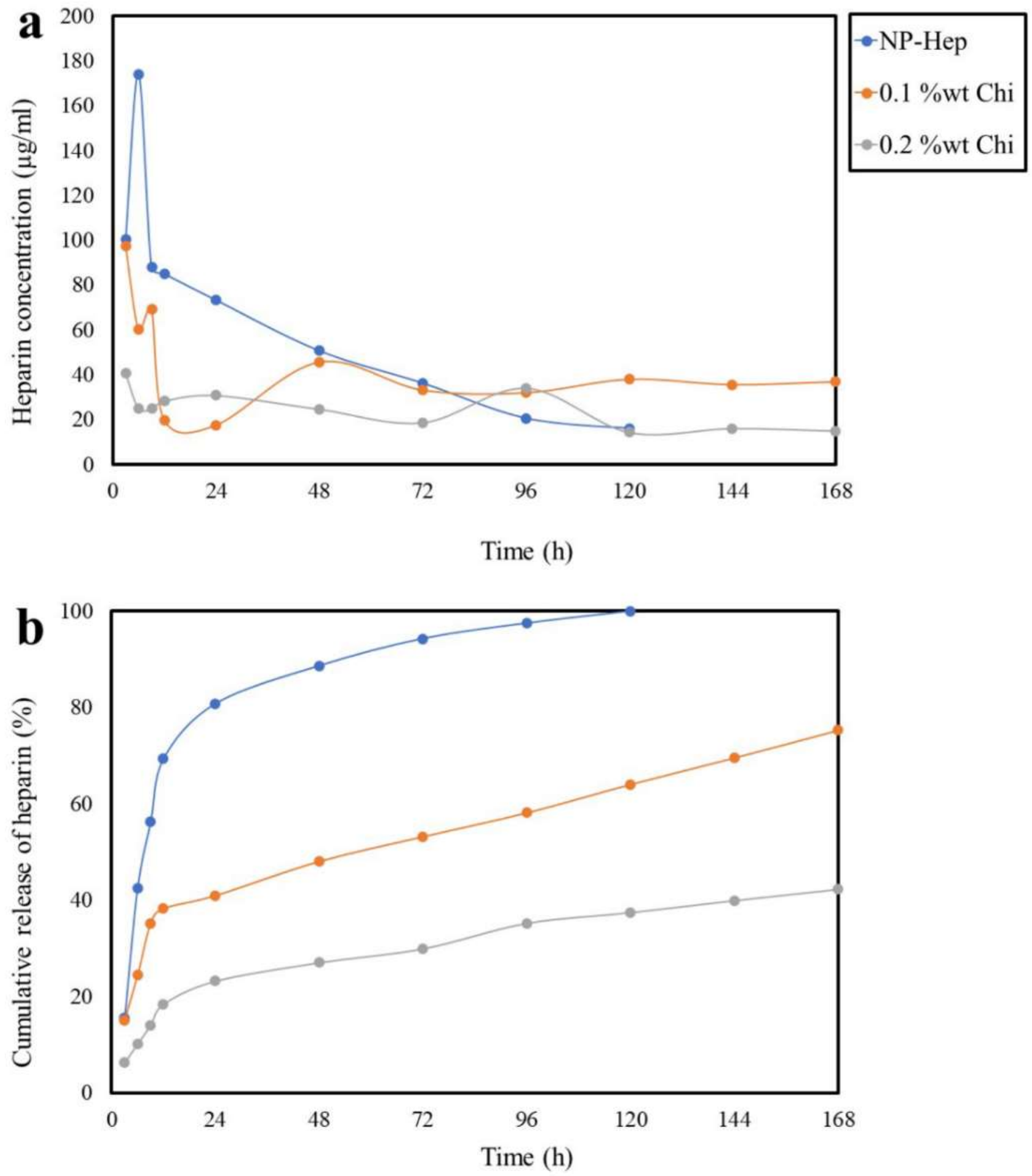

Figure 7. Graph of the release concentration (a) and cumulative level (b) of heparin from the samples.

Regarding the drug delivery behavior of nanotubular and nanoporous platforms for hydrophilic drugs, gentamicin-loaded Ti-O nanotubes with the average diameter of 120 nm and the thickness of 50 µm have shown a total release within 4 days (Gulati et al., 2012). On the contrary, the drug release of 100% from the NP-Hep sample produced in this work was obtained in 5 days. The slower drug release from our sample is owing to the smaller diameter of the nanopores (about 45 nm), which affects the amount of PBS diffusion inside the porous layer. Also, in studies conducted on the release of vancomycin from Ti-O nanotubes with an approximate thickness of 50 μm and diameter of 30-50 nm, 100% of the drug was released within 6-12 h (Hamlekhan et al., 2015; Zhang et al., 2013). This result reflects the slow release from our anodized sample with the thickness of about 8.1 ± 1.8 μm. The most important reason for this difference is related to the drug loading method used. In this study, a vacuum setup was used to increase the drug loading capacity inside the nanopores and to slow down the drug release kinetics from the structure. In conventional loading techniques like gravity immersion and surface absorption, capillary forces originating from the small dimensions of nanopores as well as air molecules located above the nanopores in the course of the drug impregnation process restrict the level and depth of the drug diffusion into nanopores (Aw et al., 2011). Using a vacuum setup not only limits the number of air molecules that prevent the complete diffusion of the drug into the pores, but also creates an additional force for the better diffusion of the drug in the depth of the pores, increasing the drug loading capacity. Another parameter that is responsible for an enhanced drug loading level is the low contact angle achieved after the anodizing process. Heparin is hydrophilic and water-soluble; that is, the low contact angle of the surface can facilitate the loading of heparin into the pores.

The burst release is significantly controlled using the chitosan coatings, where the accumulative release in the first 12 h is almost 38 and 18 % from the 0.1 and 0.2 wt% Chi samples, respectively. In the sustained release stage, after 120 h of soaking the samples in the PBS solution, 100% of heparin is released from the NP-Hep sample. But the application of the chitosan coatings causes a more stable release of heparin and after 7 days, 75 and 40% of heparin are released from the samples with a thin and thick coatings, respectively. Chitosan is a biopolymer with molecular permeability and degradability, allowing drug molecules to penetrate through the polymer coating applied on the nanoporous layer. As the chitosan layers act as a physical barrier, the level of the drug released from such a system is reversely proportional to the thickness of the applied polymer layer; that is, the thicker the layer, the slower the release.

Comparatively, in a study done on a three-layer composite made of hydroxyapatite/calcium phosphate-heparin/gelatin-heparin, over 35% of heparin was released within the first 12 h (Lai et al., 2018). In another study conducted on the release of heparin from INP hydrogels, approximately 33% of heparin was released within the first 12 h (Gritsch et al., 2018). Nevertheless, the best sample of the present study (0.2 wt% Chi) releases 18 % of heparin in the first 12 h, which is much less than the amounts mentioned in the above-mentioned studies. According to the literature (Wei et al., 2013), the release of at least 1.7 $\mu g/cm^2h$ of heparin provides no clot formation, i.e., proper blood compatibility. It also has a positive effect on the growth of endothelial cells and the suppression of muscle cells. The amount of cumulative release up to 600 μg/ml does not impose any toxicity on hemocompatibility, endothelial cells, and muscle cells (Fellows et al., 2018; Khoshnood et al., 2017; Mohammadi et al., 2019b; Qiu et al., 2019). Since the level of heparin released from the samples coated with chitosan is in this range, using such local drug delivery systems seems to be promising from the controlled release viewpoint due to the positive role of nanopores and chitosan capsulates.

To further investigate the drug delivery behavior from the drug-loaded samples, the associated experimental results were fitted to the Higuchi (Dimakos et al., 2016), zero-order (Chu et al., 2008), and first-order (Rahimipour et al., 2020b) models in both burst and sustained release stages. According to Table 1 that lists the correlation coefficient of the regressions (Rc), the NP-Hep sample during the first 12 h has the highest match with the first-order model and with the Higuchi model during the sustained release mode. The former confirms a fast, concentration-dependent drug release realized from the first-order model (Davoodian et al., 2020; Siepmann & Siepmann, 2012) and the latter reflects the control of the drug release by a diffusional process based on the Higuchi model (Siepmann & Siepmann, 2012). The relatively low agreement of the NP-Hep sample with the zero-order model is as a result of the larger diameter of nanopores (44.3 nm) compared to heparin molecules (~9 nm).

Table 1. Rc value of the NP-Hep, 0.1 wt% Chi and 0.2 wt% Chi samples for fitting with the different kinetic models

| **Model** | **Rc** | | | | | |
|---|---|---|---|---|---|---|
| | **NP-Hep** | | **0.1 wt% Chi** | | **0.2 wt% Chi** | |
| | **Burst phase** | **Sustained phase** | **Burst phase** | **Sustained phase** | **Burst phase** | **Sustained phase** |
| **Higuchi** | 0.92 | 0.95 | 0.98 | 0.98 | 0.95 | 0.99 |
| **First-order** | 0.98 | 0.99 | 0.91 | 0.95 | 0.99 | 0.98 |
| **Zero-order** | 0.96 | 0.88 | 0.98 | 0.98 | 0.99 | 0.97 |

According to Table 1, the heparin release parameters from the chitosan-coated samples show an acceptable agreement with the Higuchi model for the rapid and sustained release modes. Compared to the NP-Hep sample, the transition of the first-order mode to the diffusional Higuchi behavior for the early stage of release indicates a shortening of the fast release owing to the application of the chitosan coating. The agreement of the sustained release phase for the chitosan-coated sample with the Higuchi model also indicates the control of heparin delivery in this phase by diffusion.

The Korsmeyer-Peppas model (Rahimipour et al., 2020b) was also used for the 0.1 and 0.2 wt% Chi samples.

$$F = \left(\frac{M_t}{M_\infty}\right) = K_m t^n \tag{5}$$

where $M_t$ is the content of the drug released at period $t$, $M_\infty$ is the entire level of the drug, $K_m$ and $n$ are the kinetic value and delivery exponent, respectively (Mohammadi et al., 2019b). The parameter $n$ in this equation can describe various delivery mechanisms. $n < 0.5$ shows a diffusion behavior, whereas $n > 1$ means a degradation behavior. Amounts between these limits suggest the presence of both mechanisms (Davoodian et al., 2020). This value for the initial burst release stage for 0.1 and 0.2 wt% Chi was calculated to be 0.67 and 0.8, respectively, which indicates the involvement of the two mechanisms of degradation and diffusion. But this value for the subsequent sustained phase for 0.1 and 0.2 wt% Chi was 0.33 and 0.32, respectively, which suggests the control of the drug delivery by a diffusion mechanism.

### *3.4. Cell cytocompatibility studies*

The results of the HUVEC viability on the various specimens in 1 and 3 days of culture are depicted in Figure 8. On the first day of culture, there is a considerable difference only between the MP-NiTi and NP-NiTi samples. It is related to the impact of surface morphology and wettability on the cell viability and adhesion. Despite the enhanced wettability of the drug-loaded samples (NP-Hep, 0.1 and 0.2 wt% Chi), the insignificant difference between the MP-NiTi and drug-coated samples can be associated with the negative effect of heparin. It is also noticeable that the chitosan-coated samples do not benefit from nanorough surfaces.

Surface roughness and wettability are the major factors influencing the attachment and growth of endothelial cells on surfaces (Ding et al., 2015; Gong et al., 2019a). Nanotubular, nanoporous, grooved and other nanoscale patterned surfaces primarily alter the attachment and growth of endothelial cells compared to smooth surfaces, where these features are rarely affected by the surface chemistry on the early time of culture. In addition, the anodic layer of titanium and its alloys has a negative charge, which can alter cell adhesion and proliferation. Because cells adhere to these surfaces via electrostatic interactions, giving rise to the enhanced cell proliferation (Gong et al., 2019b). Regarding wettability, hydrophilic surfaces can adsorb high amounts of water molecules. Accordingly, they can alter cell activities through the modification and adsorption of extracellular matrix proteins. In addition, hydrophilic surfaces can bond with cerium proteins through electrostatic attractions or hydrogen bonds, providing a high density of available molecules attached to the surface for cell attachment and growth.

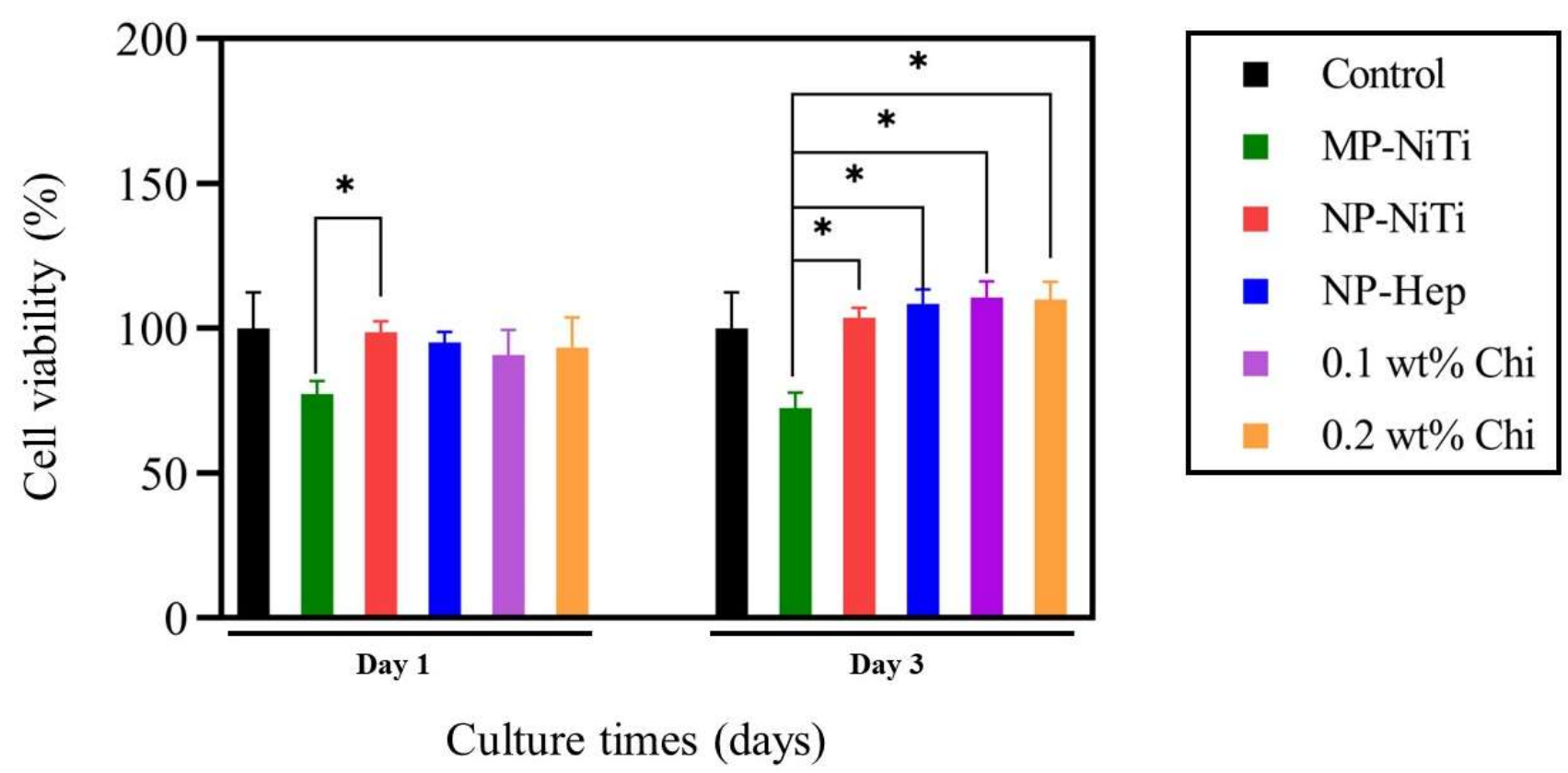

Figure 8. MTT assay results on the different samples.

On the third day, there is a considerable difference between the MP-NiTi and other samples, but there is no noteworthy difference between the heparin-loaded samples. The release of ions from MP-NiTi into the environment is responsible for its lower cell cytocompatibility, which is controlled by anodizing and chitosan coating in the other samples. The minor improvement in the cell viability of the chitosan-coated samples in comparison to the NiTi-Hep sample can be related to the effect of surface chemistry (chitosan) and controlled delivery of the drug and ions. According to the literature (Ding et al., 2015; Liu et al., 2014b;

Meng et al., 2009; Yao et al., 2014), chitosan accelerates the reendothelialization process because it improves epithelial cell adhesion and growth by stimulating fibroblasts to release interleukin-8. Also, heparin at the optimal levels is effective in the growth of cells and causes their recovery. The negative charge of heparin creates an electrostatic attraction with endothelial cells with a positive charge. Therefore, the combined use of chitosan and heparin at the optimal doses has improved the proliferation of endothelial cells. Considering the similar cell viability results of the 0.1 and 0.2 wt% Chi samples and the short half-life of heparin (1 to 2 h (X. Yang et al., 2022)), 0.2 wt% Chi can be regarded as the optimal sample because its release profile of heparin is longer and can cover the short half-life of heparin for a longer period.

## 4. Conclusion

In this study, a regular Ni-Ti-O nanoporous layer was fabricated on nitinol by a two-stage anodization process for heparin loading and delivery. Chitosan coatings were also used to control the release of heparin and alter the surface chemistry. The evaluation of wettability as an effective parameter in biological responses showed that the effect of the surface roughening caused by anodization prevails over that of the chitosan coating. The samples coated with chitosan presented a controlled drug delivery kinetics compared to the uncoated sample, improving the metabolism of endothelial cells. Also, it was found that the grown nanopores provided a more loading capacity and controlled delivery of heparin in comparison to conventional anodically-grown nanopores. Because the systems developed in the work benefited from a regular nanoporous layer and a vacuum setup for the drug loading, followed by a further controlled delivery behavior after chitosan coating within the therapeutic window.

## Acknowledgment

Part of the research reported in this paper was supported by National Institute of Dental & Craniofacial Research of the National Institutes of Health under award number R15DE027533, R56 DE029191 and 3R15DE027533-01A1W1. The content is solely the responsibility of the authors and does not necessarily represent the official views of the National Institutes of Health.

## References

Aw, M. S., Gulati, K., & Losic, D. (2011). Controlling Drug Release from Titania Nanotube Arrays Using Polymer Nanocarriers and Biopolymer Coating. *Journal of Biomaterials and Nanobiotechnology*, *02*(05), 477–484. https://doi.org/10.4236/JBNB.2011.225058

Bartorelli, A. L., Trabattoni, D., Fabbiocchi, F., Montorsi, P., De Martini, S., Calligaris, G., Teruzzi, G., Galli, S., & Ravagnani, P. (2003). Synergy of Passive Coating and Targeted Drug Delivery: *Journal of Interventional Cardiology*, *16*(6), 499–505. https://doi.org/10.1046/J.1540-8183.2003.01050.X

Bhargava, B., Reddy, N. K., Karthikeyan, G., Raju, R., Mishra, S., Singh, S., Waksman, R., Virmani, R., & Somaraju, B. (2006). A novel paclitaxel-eluting porous carbon–carbon nanoparticle coated, nonpolymeric cobalt–chromium stent: Evaluation in a porcine model. *Catheterization and Cardiovascular Interventions*, *67*(5), 698–702. https://doi.org/10.1002/CCD.20698

Byrne, R. S., & Deasy, P. B. (2002). Use of commercial porous ceramic particles for sustained drug delivery. *International Journal of Pharmaceutics*, *246*(1–2), 61–73. https://doi.org/10.1016/S0378-5173(02)00357-5

Chen, J., Huang, N., Li, Q., Chu, C. H., Li, J., & Maitz, M. F. (2016a). The effect of electrostatic heparin/collagen layer-by-layer coating degradation on the biocompatibility. *Applied Surface Science*, *362*, 281–289. https://doi.org/10.1016/J.APSUSC.2015.11.227

Chen, J., Huang, N., Li, Q., Chu, C. H., Li, J., & Maitz, M. F. (2016b). The effect of electrostatic heparin/collagen layer-by-layer coating degradation on the biocompatibility. *Applied Surface Science*, *362*, 281–289. https://doi.org/10.1016/J.APSUSC.2015.11.227

Chen, W., Habraken, T. C. J., Hennink, W. E., & Kok, R. J. (2015). Polymer-Free Drug-Eluting Stents: An Overview of Coating Strategies and Comparison with Polymer-Coated Drug-Eluting Stents. *Bioconjugate Chemistry*, *26*(7), 1277–1288. https://doi.org/10.1021/ACS.BIOCONJCHEM.5B00192/ASSET/IMAGES/MEDIUM/BC-2015-001929_0009.GIF

Chu, C. L., Wang, R. M., Hu, T., Yin, L. H., Pu, Y. P., Lin, P. H., Wu, S. L., Chung, C. Y., Yeung, K. W. K., & Chu, P. K. (2008). Surface structure and biomedical properties of chemically polished and electropolished NiTi shape memory alloys. *Materials Science and Engineering: C*, *28*(8), 1430–1434. https://doi.org/https://doi.org/10.1016/j.msec.2008.03.009

Davoodian, F., Salahinejad, E., Sharifi, E., Barabadi, Z., & Tayebi, L. (2020). PLGA-coated drug-loaded nanotubes anodically grown on nitinol. *Materials Science and Engineering: C*, *116*, 111174. https://doi.org/10.1016/J.MSEC.2020.111174

Deuse, T., Erben, R. G., Ikeno, F., Behnisch, B., Boeger, R., Connolly, A. J., Reichenspurner, H., Bergow, C., Pelletier, M. P., Robbins, R. C., & Schrepfer, S. (2008). Introducing the first polymer-free leflunomide eluting stent. *Atherosclerosis*, *200*(1), 126–134. https://doi.org/10.1016/J.ATHEROSCLEROSIS.2007.12.055

Dimakos, K., Mariotto, A., & Giacosa, F. (2016). Optimization Of The Fatigue Resistance Of Nitinol Stents Through Shot Peening. *Procedia Structural Integrity*, *2*, 1522–1529. https://doi.org/https://doi.org/10.1016/j.prostr.2016.06.193

Ding, Y., Yang, M., Yang, Z., Luo, R., Lu, X., Huang, N., Huang, P., & Leng, Y. (2015). Cooperative control of blood compatibility and re-endothelialization by immobilized heparin and substrate topography. *Acta Biomaterialia*, *15*, 150–163. https://doi.org/10.1016/J.ACTBIO.2014.12.014

Drugacz, J., Lekston, Z., Morawiec, H., & Januszewski, K. (1995). Use of TiNiCo shape-memory clamps in the surgical treatment of mandibular fractures. *Journal of Oral and Maxillofacial Surgery*, *53*(6), 665–671. https://doi.org/10.1016/0278-2391(95)90166-3

Fellows, B., Ghobrial, N., Mappus, E., … A. H.-, Medicine, B. and, & 2018, undefined. (n.d.). In vitro studies of heparin-coated magnetic nanoparticles for use in the treatment of neointimal hyperplasia. *Elsevier*. Retrieved April 10, 2023, from https://www.sciencedirect.com/science/article/pii/S1549963418300583

Gong, Z., Hu, Y., Gao, F., Quan, L., Liu, T., Gong, T., & Pan, C. (2019a). Effects of diameters and crystals of titanium dioxide nanotube arrays on blood compatibility and endothelial cell behaviors. *Colloids and Surfaces B: Biointerfaces*, *184*, 110521. https://doi.org/10.1016/J.COLSURFB.2019.110521

Gong, Z., Hu, Y., Gao, F., Quan, L., Liu, T., Gong, T., & Pan, C. (2019b). Effects of diameters and crystals of titanium dioxide nanotube arrays on blood compatibility and endothelial cell behaviors. *Colloids and Surfaces B: Biointerfaces*, *184*, 110521. https://doi.org/10.1016/J.COLSURFB.2019.110521

Gritsch, L., Motta, F. L., Contessi Negrini, N., Yahia, L. H., & Farè, S. (2018). Crosslinked gelatin hydrogels as carriers for controlled heparin release. *Materials Letters*, *228*, 375–378. https://doi.org/10.1016/J.MATLET.2018.06.047

Gulati, K., Ramakrishnan, S., Aw, M. S., Atkins, G. J., Findlay, D. M., & Losic, D. (2012). Biocompatible polymer coating of titania nanotube arrays for improved drug elution and osteoblast adhesion. *Acta Biomaterialia*, *8*(1), 449–456. https://doi.org/10.1016/J.ACTBIO.2011.09.004

Hamlekhan, A., Sinha-Ray, S., Takoudis, C., Mathew, M. T., Sukotjo, C., Yarin, A. L., & Shokuhfar, T. (2015). Fabrication of drug eluting implants: study of drug release mechanism from titanium dioxide nanotubes. *Journal of Physics D: Applied Physics*, *48*(27), 275401. https://doi.org/10.1088/0022-3727/48/27/275401

Hang, R., Huang, X., Tian, L., He, Z., & Tang, B. (2012). Preparation, characterization, corrosion behavior and bioactivity of Ni2O3-doped TiO2 nanotubes on NiTi alloy. *Electrochimica Acta*, *70*, 382–393. https://doi.org/10.1016/J.ELECTACTA.2012.03.085

Hang, R., Liu, Y., Bai, L., Zhang, X., Huang, X., Jia, H., & Tang, B. (2018). Length-dependent corrosion behavior, Ni2+ release, cytocompatibility, and antibacterial ability of Ni-Ti-O nanopores anodically grown on biomedical NiTi alloy. *Materials Science and Engineering: C*, *89*, 1–7. https://doi.org/10.1016/J.MSEC.2018.03.018

Hang, R., Liu, Y., Bai, L., Zong, M., Wang, X., Zhang, X., Huang, X., & Tang, B. (2017). Electrochemical synthesis, corrosion behavior and cytocompatibility of Ni-Ti-O nanopores on NiTi alloy. *Materials Letters*, *202*, 5–8. https://doi.org/10.1016/J.MATLET.2017.05.089

Hang, R., Liu, Y., Gao, A., Zong, M., Bai, L., … X. Z.-S. and C., & 2017, undefined. (n.d.). Fabrication of Ni-Ti-O nanoporous film on NiTi alloy in ethylene glycol containing NaCl. *Elsevier*. Retrieved February 28, 2023, from https://www.sciencedirect.com/science/article/pii/S025789721730381X

Hang, R., Zhao, Y., Bai, L., Liu, Y., Gao, A., Zhang, X., Huang, X., Tang, B., & Chu, P. K. (2017). Fabrication of irregular-layer-free and diameter-tunable Ni–Ti–O nanopores by anodization of NiTi alloy. *Elsevier*. https://doi.org/10.1016/j.elecom.2017.01.010

Huang, J., Dong, P., Hao, W., Wang, T., Xia, Y., Da, G., & Fan, Y. (2014). Biocompatibility of TiO2 and TiO2/heparin coatings on NiTi alloy. *Applied Surface Science*, *313*, 172–182. https://doi.org/10.1016/J.APSUSC.2014.05.182

Huang, Q., Yang, Y., Zheng, D., Song, R., Zhang, Y., Jiang, P., Vogler, E. A., & Lin, C. (2017). Effect of construction of TiO2 nanotubes on platelet behaviors: Structure-property relationships. *Acta Biomaterialia*, *51*, 505–512. https://doi.org/10.1016/J.ACTBIO.2017.01.044

Jarosz, M., Pawlik, A., Szuwarzyński, M., Jaskuła, M., & Sulka, G. D. (2016). Nanoporous anodic titanium dioxide layers as potential drug delivery systems: Drug release kinetics and mechanism. *Colloids and Surfaces B: Biointerfaces*, *143*, 447–454. https://doi.org/10.1016/J.COLSURFB.2016.03.073

Khoshnood, N., Zamanian, A., & Massoudi, A. (2017). Tailoring in vitro drug delivery properties of titania nanotubes functionalized with (3-Glycidoxypropyl) trimethoxysilane.

*Materials Chemistry and Physics*, *193*, 290–297. https://doi.org/10.1016/J.MATCHEMPHYS.2017.02.044

Lai, Y., Cheng, P., Yang, C., Films, S. Y.-T. S., & 2018, undefined. (n.d.). Electrolytic deposition of hydroxyapatite/calcium phosphate-heparin/gelatin-heparin tri-layer composites on NiTi alloy to enhance drug loading and prolong releasing . *Elsevier*. Retrieved April 10, 2023, from https://www.sciencedirect.com/science/article/pii/S0040609018300671

Liu, T., Zeng, Z., Liu, Y., Wang, J., Maitz, M. F., Wang, Y., Liu, S., Chen, J., & Huang, N. (2014a). Surface modification with dopamine and heparin/poly-l-lysine nanoparticles provides a favorable release behavior for the healing of vascular stent lesions. *ACS Applied Materials and Interfaces*, *6*(11), 8729–8743. https://doi.org/10.1021/AM5015309/ASSET/IMAGES/MEDIUM/AM-2014-015309_0009.GIF

Liu, T., Zeng, Z., Liu, Y., Wang, J., Maitz, M. F., Wang, Y., Liu, S., Chen, J., & Huang, N. (2014b). Surface Modification with Dopamine and Heparin/Poly-l-Lysine Nanoparticles Provides a Favorable Release Behavior for the Healing of Vascular Stent Lesions. *ACS Applied Materials & Interfaces*, *6*(11), 8729–8743. https://doi.org/10.1021/am5015309

Mani, G., Johnson, D. M., Marton, D., Feldman, M. D., Patel, D., Ayon, A. A., & Agrawal, C. M. (2008). Drug delivery from gold and titanium surfaces using self-assembled monolayers. *Biomaterials*, *29*(34), 4561–4573. https://doi.org/10.1016/J.BIOMATERIALS.2008.08.014

Meng, S., Liu, Z., Shen, L., Guo, Z., Chou, L. L., Zhong, W., Du, Q., & Ge, J. (2009). The effect of a layer-by-layer chitosan–heparin coating on the endothelialization and coagulation properties of a coronary stent system. *Biomaterials*, *30*(12), 2276–2283. https://doi.org/10.1016/J.BIOMATERIALS.2008.12.075

Mohammadi, F., Golafshan, N., Kharaziha, M., & Ashrafi, A. (2019a). Chitosan-heparin nanoparticle coating on anodized NiTi for improvement of blood compatibility and biocompatibility. *International Journal of Biological Macromolecules*, *127*, 159–168. https://doi.org/10.1016/J.IJBIOMAC.2019.01.026

Mohammadi, F., Golafshan, N., Kharaziha, M., & Ashrafi, A. (2019b). Chitosan-heparin nanoparticle coating on anodized NiTi for improvement of blood compatibility and biocompatibility. *International Journal of Biological Macromolecules*, *127*, 159–168. https://doi.org/10.1016/J.IJBIOMAC.2019.01.026

Mousavi, S. A., Moshfeghi, A., Davoodian, F., & Salahinejad, E. (2021). Eliminating the irregular surface layer of anodically-grown Ni-Ti-O nanopore arrays in a two-stage anodization. *Surface and Coatings Technology*, *405*, 126707. https://doi.org/10.1016/J.SURFCOAT.2020.126707

Oh, K. T., Choo, S. U., Kim, K. M., & Kim, K. N. (2005). A stainless steel bracket for orthodontic application. *European Journal of Orthodontics*, *27*(3), 237–244. https://doi.org/10.1093/EJO/CJI005

Pan, C. J., Pang, L. Q., Gao, F., Wang, Y. N., Liu, T., Ye, W., & Hou, Y. H. (2016). Anticoagulation and endothelial cell behaviors of heparin-loaded graphene oxide coating on titanium surface. *Materials Science and Engineering: C*, *63*, 333–340. https://doi.org/10.1016/J.MSEC.2016.03.001

Peters, S., Behnisch, B., Heilmann, T., & Richter, C. (2009). First-in-man use of polymer-free valsartan-eluting stents in small coronary vessels: a comparison to polymer-free rapamycin (2%)-eluting stents. *Http://Dx.Doi.Org/10.1177/1470320308098591*, *10*(2), 91–95. https://doi.org/10.1177/1470320308098591

Qiu, H., Qi, P., Liu, J., Yang, Y., Tan, X., Xiao, Y., Biomaterials, M. M.-, & 2019, undefined. (n.d.). Biomimetic engineering endothelium-like coating on cardiovascular stent through heparin and nitric oxide-generating compound synergistic modification strategy. *Elsevier*. Retrieved April 10, 2023, from https://www.sciencedirect.com/science/article/pii/S0142961219301838

Rahimipour, S., Salahinejad, E., Sharifi, E., Nosrati, H., & Tayebi, L. (2020a). Structure, wettability, corrosion and biocompatibility of nitinol treated by alkaline hydrothermal and hydrophobic functionalization for cardiovascular applications. *Applied Surface Science*, *506*. https://doi.org/10.1016/J.APSUSC.2019.144657

Rahimipour, S., Salahinejad, E., Sharifi, E., Nosrati, H., & Tayebi, L. (2020b). Structure, wettability, corrosion and biocompatibility of nitinol treated by alkaline hydrothermal and hydrophobic functionalization for cardiovascular applications. *Applied Surface Science*, *506*, 144657. https://doi.org/10.1016/J.APSUSC.2019.144657

Shen, Y., Tang, C., Sun, B., Zhang, Y., … X. S.-C. E., & 2022, undefined. (n.d.). 3D printed personalized, heparinized and biodegradable coronary artery stents for rabbit abdominal aorta implantation. *Elsevier*. Retrieved April 10, 2023, from https://www.sciencedirect.com/science/article/pii/S1385894722036865

Siepmann, J., & Siepmann, F. (2012). Modeling of diffusion controlled drug delivery. *Journal of Controlled Release*, *161*(2), 351–362. https://doi.org/10.1016/J.JCONREL.2011.10.006

Tada, N., Virmani, R., Grant, G., Bartlett, L., Black, A., Clavijo, C., Christians, U., Betts, R., Savage, D., Su, S. H., Shulze, J., & Kar, S. (2010). Polymer-Free Biolimus A9-Coated Stent Demonstrates More Sustained Intimal Inhibition, Improved Healing, and Reduced Inflammation Compared With a Polymer-Coated Sirolimus-Eluting Cypher Stent in a Porcine Model. *Circulation: Cardiovascular Interventions*, *3*(2), 174–183. https://doi.org/10.1161/CIRCINTERVENTIONS.109.877522

Thipparaboina, R., Khan, W., & Domb, A. J. (2013). Eluting combination drugs from stents. *International Journal of Pharmaceutics*, *454*(1), 4–10. https://doi.org/10.1016/J.IJPHARM.2013.07.005

Watt, J., Kennedy, S., McCormick, C., Agbani, E. O., McPhaden, A., Mullen, A., Czudaj, P., Behnisch, B., Wadsworth, R. M., & Oldroyd, K. G. (2013). Succinobucol-eluting stents increase neointimal thickening and peri-strut inflammation in a porcine coronary model. *Catheterization and Cardiovascular Interventions*, *81*(4), 698–708. https://doi.org/10.1002/CCD.24473

Wei, H., Han, L., Ren, J., & Jia, L. (2013). Anticoagulant surface coating using composite polysaccharides with embedded heparin-releasing mesoporous silica. *ACS Applied Materials and Interfaces*, *5*(23), 12571–12578. https://doi.org/10.1021/AM403882X

Yang, X., Wang, Q., Zhang, A., Shao, X., Liu, T., Tang, B., & Fang, G. (2022). Strategies for sustained release of heparin: A review. *Carbohydrate Polymers*, *294*, 119793. https://doi.org/10.1016/J.CARBPOL.2022.119793

Yang, Z., Tu, Q., Wang, J., & Huang, N. (2012). The role of heparin binding surfaces in the direction of endothelial and smooth muscle cell fate and re-endothelialization. *Biomaterials*, *33*(28), 6615–6625. https://doi.org/10.1016/J.BIOMATERIALS.2012.06.055

Yao, Y., Wang, J., Cui, Y., Xu, R., Wang, Z., Zhang, J., Wang, K., Li, Y., Zhao, Q., & Kong, D. (2014). Effect of sustained heparin release from PCL/chitosan hybrid small-diameter vascular grafts on anti-thrombogenic property and endothelialization. *Acta Biomaterialia*, *10*(6), 2739–2749. https://doi.org/10.1016/J.ACTBIO.2014.02.042

Yu, C., Yang, H., Wang, L., Thomson, J., … L. T.-M. S. and, & 2021, undefined. (n.d.). Surface modification of polytetrafluoroethylene (PTFE) with a heparin-immobilized extracellular matrix (ECM) coating for small-diameter vascular grafts applications. *Elsevier*. Retrieved April 10, 2023, from https://www.sciencedirect.com/science/article/pii/S0928493121004409

Zhang, H. Z., Sun, Y., Tian, A., Xue, X. X., Wang, L., Alquhali, A., & Bai, X. Z. (2013). Improved antibacterial activity and biocompatibility on vancomycin-loaded TiO2 nanotubes: in vivo and in vitro studies. *International Journal of Nanomedicine*, *8*, 4379. https://doi.org/10.2147/IJN.S53221

Zhao, Y., Liu, Y., Liu, S., Bai, L., Yao, X., Tang, B., & Hang, R. (2019). THE INFLUENCE of ELECTROLYTE STIRRING on ANODIC GROWTH of Ni-Ti-O NANOPORES on NiTi ALLOY. *Surface Review and Letters*, *26*(3). https://doi.org/10.1142/S0218625X18501627